\documentclass[useAMS,usenatbib]{mn2e}
\usepackage{times}
\usepackage{graphicx, latexsym, amssymb, amscd, psfrag}
\usepackage{epsfig}
\def\eg{{e.g.}}
\def\ie{{i.e.}}
\def\ssf{{log SFR/M$^*$}}
\def\simgt{\hbox{\rlap{\raise 0.425ex\hbox{$>$}}\lower 0.65ex\hbox{$\sim$}}}
\def\simlt{\hbox{\rlap{\raise 0.425ex\hbox{$<$}}\lower 0.65ex\hbox{$\sim$}}}
\title[Red star-forming and blue passive galaxies in clusters]{Red star-forming
 and blue passive galaxies in clusters}
\author[Mahajan and Raychaudhury]{Smriti Mahajan\thanks{E-mail:
sm@star.sr.bham.ac.uk} and Somak Raychaudhury \\  
School of Physics and Astronomy, College of Engineering and Physical Sciences, 
University of Birmingham, Birmingham B15~2TT, United Kingdom}

\begin{document}

\date{}

\pagerange{\pageref{firstpage}--\pageref{lastpage}} \pubyear{2009}
\maketitle

\label{firstpage}
%======================== ABSTRACT ==================================
\begin{abstract}
  We explore the relation between colour (measured from photometry)
  and specific star formation rate (derived from optical spectra
  obtained by SDSS DR4) of over 6000 galaxies ($M_r\!\le\! -20.5$) in
  and around ($<3\,r_{200}$) low redshift ($z<0.12$) Abell
  clusters. Even though, as expected, most red sequence galaxies have
  little or no ongoing star formation, and most blue galaxies are
  currently forming stars, there are significant populations of red
  star-forming and blue passive galaxies. This paper examines various
  properties of galaxies belonging to the latter two categories, to
  understand why they deviate from the norm. These properties include
  morphological parameters, internal extinction, spectral features
  such as EW(H$_\delta$) and the 4000~\AA\ break, and metallicity. Our
  analysis shows that the blue passive galaxies have properties very
  similar to their star-forming counterparts, except that their large
  range in H$_\delta$ equivalent width indicates recent truncation of
  star formation. The red star-forming galaxies fall into two broad
  categories, one of them being massive galaxies in cluster cores
  dominated by an old stellar population, but with evidence of current
  star formation in the core (possibly linked with AGN). For the
  remaining red star-forming galaxies it is evident from spectral
  indices, stellar and gas-phase metallicities and mean stellar ages
  that their colours result from the predominance of a metal-rich
  stellar population. Only half of the red star-forming galaxies have
  extinction values consistent with a significant presence of
  dust. The implication of the properties of these star-forming
  galaxies on environmental studies, like that of the Butcher-Oemler
  effect, is discussed.
 \end{abstract}


%======================== Introduction===============================
\section{Introduction}

If, as is widely believed, large-scale structures form in the Universe
by means of the hierarchical assembly of dark haloes, one would expect
the most massive galaxies to be the youngest, as they would be formed
from mergers over time. However, the brightest galaxies, elliptical in
morphology, typically found at the centres of rich clusters, appear to
be dominated by old stellar populations. Galaxies with current star
formation, on the other hand, have been shown to be disc-dominated,
mostly occurring in regions of lower density. Such relations between
the properties of galaxies and their environment were first pointed out
using morphology to characterise galaxies \citep{o74,ms,d2}, but, in
later studies, other photometric properties such as colour
\citep[e.g.][]{kauff03a,b6,w1,cdeep} have been shown to correlate with
environment as well.

In addition to understanding the formation of dark matter haloes, the
study of galaxy formation and evolution clearly requires detailed
knowledge of the various physical processes that affect baryons, such
as mergers, tidal interactions, stripping, cooling and feedback.
Observations of the relevant galaxy features and their dependence on
environment are very important in our comprehension of the relative
importance of these processes, a subject that remains an issue of
considerable disagreement \citep[see, e.g.,][]{oem09}.

The availability of large-scale spectroscopic surveys, such as the
Sloan Digital Sky survey (henceforth SDSS), has resulted in structural
parameters of galaxies being largely replaced by spectroscopic
properties, in the exploration of the impact of the global and local
environment on the nature of a galaxy \citep[][among
others]{b7,h2,hb,pk06,p1,w1,p2,p08,mrp09}. Compared to broad-band
colours, spectroscopic parameters such as integrated star formation
rate (SFR), specific star formation rate (SSFR or SFR/M$^*$),
H$_\delta$ equivalent width (EW) and the 4000 \AA\ break (D$_n$4000), 
in principle, give
more detailed information about the underlying stellar populations in
galaxies, but one cannot expect these parameters to behave in the same
way with environment as the global photometric properties.

Various factors, including the presence of dust, and the well-known
degeneracy between the effect of age and metallicity, prevent a
straight-forward correspondence between the star formation history of
a galaxy and its broad-band colours \citep[e.g.,][]{faber,cal97}.  As
a result, the use of broad-band colours and of spectroscopic
indicators might yield different classification for a galaxy on an
evolutionary sequence. In order to compare observational studies where
different photometric or spectroscopic parameters have been used to
define sub-samples, one would like to ask, for instance, for what
fraction of galaxies the correspondence between morphology, colour and
star formation properties fails to hold, and whether such galaxies
could be identified from other observable properties. Are these
unusual galaxies more likely to occur in certain kinds of environment?

In order to address some of these questions, we choose, for this study,
galaxies in and around rich clusters in the low redshift
Universe. Galaxy clusters are well suited for such an analysis, since
from their cores to their infall regions, they provide a wide range of
environments, with varying space density and velocity dispersion. 
The aim of this work is to examine how well galaxy properties obtained
from two mutually independent techniques, namely photometry and
spectroscopy, correlate with each other, and to explore the possible
causes of discrepancy.

 The outline of this paper is as follows: in the next section we define
 the galaxy sample, and the galaxy properties to be used here.  We
 plot the specific star formation rate of these galaxies against
 colour, and identify the four classes of galaxies that we will deal
 with: red sequence, red star-forming, blue star-forming and blue
 passive.  In \S\ref{results}, we examine the various galaxy
 properties, derived from photometry and spectra, that we use in this
 analysis. The insight these properties provide into the nature of
 blue passive and red star-forming galaxies are discussed in
 \S\ref{discussion}, and the main conclusions are summarised in
 \S\ref{conclusions}. Throughout this work we adopt a cosmology with
 $\Omega_{m}=1$ \& $\Omega_{\Lambda}=0$ and $h=70$ km~s$^{-1}$
 Mpc$^{-1}$ for calculating distances and absolute magnitudes. We note
 that in the redshift range chosen for this work ($0.02\leq z \leq
 0.12$), the results are insensitive to the choice of cosmology.
 
%%%%%%%%%%%%%%%%%%%%%%%%%%%%%%%%%%%%%%%%%%%%%%%%%%%%%%%%%%%%%%%%%%%%%%%%%%%%%%%%%%%%%%%%%%%%%%%%%%%%%%%%

\section{The observational data}
\label{data}
We use the spectroscopic galaxy data provided by the SDSS
DR$4$~\citep{a2}. The SDSS consists of an imaging survey of $\pi$
steradians, mainly in the northern sky, in five passbands $u,g,r,i$ and
$z$. The imaging is done in drift-scan mode, and the data is processed
using the photometric pipeline PHOTO \citep{l1} specially written for
SDSS. The spectra are obtained using two fibre-fed double
spectrographs, covering a wavelength range of 3800-9200 \AA. The
resolution $\Delta \lambda/\lambda$ varies between 1850 and 2200.

We use the galaxy magnitudes and the corresponding galactic extinction
values in the $g$ \& $r$ SDSS passbands from the New York University
Value added galaxy Catalogue \citep[NYU-VAGC;][]{b5}. We $k$-correct
these magnitudes to $z\!=\! 0.1$ (median redshift of the sample,
\citealt{y1}). We only consider galaxies with absolute
magnitude M$_{r} \!\leq\! -20.5$ (this corresponds to the apparent
magnitude limit for the SDSS spectroscopic catalogue at the median
redshift of our sample, $z\!=\! 0.1$).

\subsection{Cluster membership}
\label{member}

We take all Abell clusters \citep{a1}, in the redshift range
0.02$<z\leq$0.12, that lie sufficiently away from the survey boundary
so as to be sampled out to $\leq6\,h^{-1}_{70}$~Mpc from the cluster centre.  We
constrain our sample to moderately rich clusters by considering only
those that have at least $20$ galaxies (brighter than the magnitude
limit, M$_{r}\leq -20.5$) with SDSS spectra within $3\,h^{-1}_{70}$~Mpc 
of the cluster centre. This leaves us with a sample of $119$ clusters.  

We aim to study the effect of present and recent star formation on the
observed properties of galaxies over a wide range of environment, and
so we seek to examine galaxies not just in the cores of the clusters,
but also on the outskirts. Detailed studies of the velocity structure
in the infall regions of clusters show that the turnaround radius of
clusters extends out to almost twice the virial radius
\citep{dg97,rines06,dunner07}, so we chose to examine radial trends of
the properties of galaxies out to a projected radius of $R\!=\!
6\,h^{-1}_{70}$~Mpc of each cluster centre. Within this radius,
we pick galaxies within
the following velocity ranges about the mean redshift of the cluster:
 \begin{equation}
 \Delta z=\left\{
   \begin{array}{ll}  
            3000 \hbox{ km s$^{-1}$} &  \textrm{~~~if $R\leq 1$~Mpc}   \\  
	    2100 \hbox{ km s$^{-1}$} &  \textrm{~~~if $1<R\leq 3$~Mpc} \\
            1500 \hbox{ km s$^{-1}$} &  \textrm{~~~if $3<R\leq 6$~Mpc} \\
 \end{array}
 \right\}.
 \label{eq:selection} 
 \end{equation}
 Adopting these differential velocity slices helps in reducing
 interlopers, in the absence of a rigorous procedure for assigning
 cluster membership, and hence is well suited for a statistical study
 like this. Note that other popular methods, such as the 3-$\sigma_v$
 velocity dispersion cut, are prone to include galaxies at large
 projected radius from the cluster centre, and our velocity cuts are
 consistent with the caustics in velocity space used by the CAIRNS survey
\citep[\eg][]{rines03,rines06}.

 \subsection{Cluster parameters}
 
 We find  the velocity dispersion  ($\sigma_v$) of each  cluster, from
 the    identified   members,    using   the    bi-weight   statistics
 \citep[\textsc{ROSTAT},][]{b1}. The characteristic size  of the cluster (the
 radius at which the mean  interior over-density in a sphere of radius
 $r$  is  $200$  times  the  critical  density  of  the  Universe)  is
 calculated    using    the     relation    given    by    \citet{c1}:
 $r_{200}=\sqrt{3}\sigma_{v}/10H(z)$. Our final sample  comprises of
 $119$  Abell clusters ($z<0.12$)  in which there are 6,010 galaxies
 within $3\,r_{200}$ of the cluster centres.

\subsection{Galaxy parameters}
\label{galaxy-data}

In this work, we use various quantities derived from the SDSS optical spectra
by \citet{b2}~(B04 henceforth),
based on SDSS DR$4$ catalogue\footnote{http://www.mpa-garching.mpg.de/SDSS/DR4/Data/sfr\_catalogue.html}.
 From this and related work \citep{kauff03b}, we use specific spectral indices (emission
corrected H$_\delta$ equivalent width and [NeIII] contamination
corrected D$_n$4000).

Two of the galaxy parameters we use to begin with, namely the specific
star formation rate (SSFR, or SFR/M$^*$) and stellar mass (M$^*$), are
derived using the SDSS galaxy spectra by B04 and \citet{kauff03b} respectively.
 B04 construct a grid of $\sim 2\times10^5$ spectral synthesis models
\citep{bruzual93,charlot01} over four relevant parameters spanning a
range in metallicity ($Z$) of $-1\! <\! \log Z/Z_{\odot} \!<\! 0.6$,
ionization parameter ($U$) $-4.0\!<\! \log U \! <\! -2.0$, total dust
attenuation ($\tau_{V}$) $0.01\! <\! \tau_{V}\! <\!4.0$ and dust-to-metal-ratio
 ($\xi$) $0.1<\xi\! <\! 0.5$. The attenuation by dust is treated using a
physically motivated model of \citet{charlot00}, which provides a
consistent model for ultra-violet (UV) to infra-red (IR) emission.
 
B04 compare the modelled line-ratios for each model in their grid
directly with those in the observed spectrum, and, for each galaxy,
construct a log-likelihood function for each point on the model grid,
summed over all lines observed in the spectrum of that galaxy. It is
worth mentioning here that since they study relative line fluxes only,
their analysis is insensitive to stellar ages, star formation history
and the relative attenuation by dust in the molecular gas clouds and
the interstellar medium.
 
 \begin{figure}
 \centering{
 {\rotatebox{0}{\epsfig{file=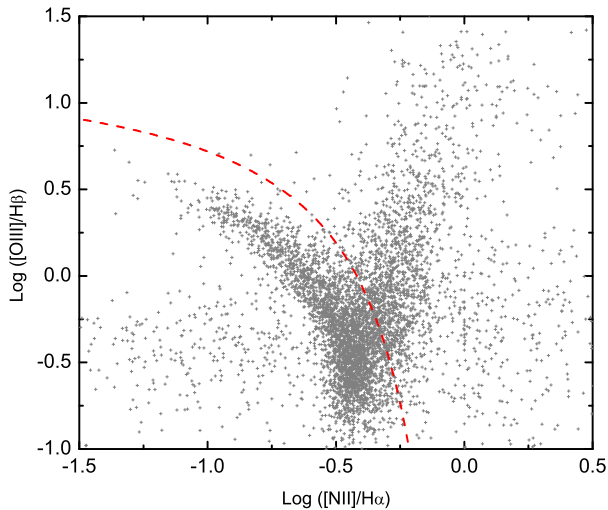,width=9.2cm}}}}
\caption{The distribution of the galaxies used in this work in the
  plot of the two line-ratios, recommended by \citet[][(BPT)]{b3},
  that are often used to discriminate between star-forming (and composite)
  galaxies (below the {\it{red}} line; \citet{kauff03c}) and the AGN-dominated
  galaxies (above). All galaxies in our sample, except the `unclassified', are
  plotted here. }
 \label{fig:bpt }
 \end{figure}

 B04 derive their values for star formation rate (SFR) and specific
 star formation rate (SFR/M$^*$) using a Bayesian technique, involving
 the likelihood functions mentioned above, using the entire spectrum
 of each galaxy (it is worth noting that the spectrum comes from a
 fibre of diameter 3$^{\prime\prime}$). It is shown that their work
 yield better estimates of the SFR and SFR/M$^*$ than any single
 spectroscopic SFR indicator, such as equivalent width (EW) of the
 H$_{\alpha}$ or O~II lines, which are used extensively in the
 literature. Furthermore, the combination of several line indices
 enables one to separate the galaxies with a significant AGN component
 from those with emission entirely due to star formation.
 The derived parameters (SFR and SFR/M$^*$) are then corrected for
 aperture biases using their photometric colours. Out of the three
 statistical estimates (mean, median and mode) for the probability
 distribution function (PDF) of SFR and
 SFR/M$^*$ derived for each galaxy, we use the median of the distribution
 in this work for each quantity, since it is independent of binning.

 Star-forming galaxies can be separated from the AGN-dominated ones on
 the basis of the so-called BPT diagram \citep{b3}, namely a plot of
 the line ratio [OIII]/H$_\beta$ against H$_\alpha$/[NII],
 Fig.~\ref{fig:bpt }, for galaxies with all four
 emission lines. 
 B04 divide all galaxies into four classes (star
 forming, AGN, composite and unclassified) using the BPT diagram. For
 those galaxies whose spectra do not show all the above mentioned emission lines,
 (called `unclassified' galaxies), and those that are classified as
 AGN-dominated, B04 adopt an indirect method for estimating the
 SFRs. They use the relationship between the D$_{n}$4000 and SFR
 derived for the high S/N ($\geq3$) star-forming galaxies (SFR$_e$) to
 estimate the SFR of the AGNs and low S/N galaxies (SFR$_{d}$; see
 section 4.1 of B04 for details).
 Using the entire spectrum also makes it possible to obtain an
 independent estimate of stellar mass M$^*$, which further constraints
 the estimated SFR/M$^*$, making it a more reliable parameter for
 evaluating star formation properties of galaxies, compared to
 other luminosity based parameters employed elsewhere
 \citep[\eg][]{lewis,g3}. 
 \citet{kauff03a} adopt similar methodology as B04 for estimating the
 stellar mass M$^*$. Their models are characterised by star-to-light ratio,
 dust attenuation and fraction of mass formed in a major burst of star formation
 (the burst mass fraction). They then compute the total mass in stars by
 multiplying the dust-corrected and $k$-corrected $z$-band luminosity of
 the galaxy, by the estimated $z$-band mass-to-light ratio of the best-fitting
 model, found using the Bayesian technique described above.
 
 Several authors have shown that the SFR of a galaxy scales with its
 stellar mass \citep[e.g.][]{noeske}. Since the reason for this
 correlation is yet unclear, and beyond the scope of this work, we
 employ the specific star formation rate SFR/M$^*$ as the more
 reliable measure of the star formation activity of galaxies.  The
 values of SFR (and SFR/M$^*$) derived for the galaxies with emission
 lines, \ie~ all but those categorised as `unclassified', are reliable
 (see Fig.~14; B04), because the classification requires S/N$>3$ in
 all the $4$ lines involved in the BPT diagram (\ie, H$_{\alpha}$,
 H$_{\beta}$, [OIII] and [NII]).  But for spectra of galaxies without
 emission lines, or of those that are unclassified (most of which are
 passive), the uncertainties are relatively higher.  We chose not to
 select galaxies based on the confidence interval values derived from
 the PDF of the SFR and SFR/M$^*$ parameters of individual galaxies,
 because excluding galaxies with higher uncertainties in SFR/M$^*$ (or
 SFR) distribution would preferentially exclude a significant number
 of passive galaxies and thus bias our samples.

 \begin{figure}
 \centering{
 {\rotatebox{270}{\epsfig{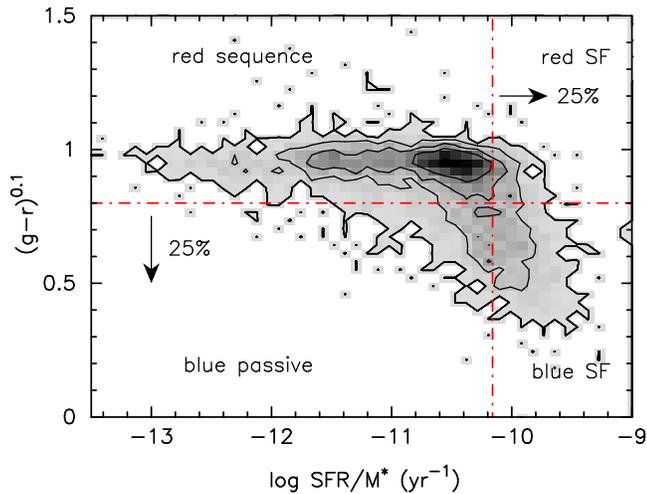}}}}
\caption{The relation between specific star formation rate (SSFR or
  SFR/$M_\ast$) for all galaxies found within $3\,r_{200}$ of the
  centre of each galaxy cluster in our sample.  The contours represent
  $150$, $100$, $50$ and $25$ galaxies respectively. The horizontal
  line marks the colour cut adopted by most Butcher-Oemler effect
  studies, \ie\ $0.2$ mag bluer than the red sequence: this labels
  $\sim25$\% of all galaxies as blue. To match this cut, we adopt a
  SSFR threshold, \ssf $= -10.16$ yr$^{-1}$ such that $25$\% of all
  the galaxies are classified as star-forming. We thus categorise our
  galaxy ensemble into four classes: ({\it{clockwise from right}})
  red star-forming, blue star-forming, blue passive and red sequence
  galaxies respectively.}
 \label{ssf-gr}
 \end{figure}

\subsection{Four categories of galaxies}
\label{sec:categories}

Galaxy properties measurable from photometry or spectra, such as
colour, SFR or SSFR (SFR/M$^*$) have been extensively used in studies
of their dependence on the environment of the galaxy
\citep[\eg][]{b7,h2,hb,pk06,w1,p2,p08,mrp09},
or on other variables representing mass, such as the stellar mass of
the galaxy \citep{kauff03b} or its circular velocity \citep{graves}.
 
In this work, we study the colour and SSFR as indicators of the
evolutionary stage of a galaxy.  In Fig.~\ref{ssf-gr} we plot the
colours of all galaxies found within $3\,r_{200}$ of the centres of
all the clusters in our sample against their specific star formation
rates.  In a pioneering study of evolution of galaxies in clusters,
\citet{bo} had adopted a colour cut to define blue galaxies as $B\!-\!
V=0.2$ mag lower than the cluster's red sequence.  Here we stack all
the cluster galaxies together, and fit a red sequence to galaxies
which has a slope of $-0.034$ with an absolute deviation of
$0.080$. Since this slope is negligible, we adopt a null slope red
sequence at $(g-r)^{0.1}\!=\! 1$. We note that using the actual red
sequence fit does not change our results at all. The blue galaxies are
thus defined to have $(g-r)^{0.1}\!\leq\! 0.8$.  This classifies
$\sim$25\% of all galaxies as blue. 

Correspondingly, we choose the SSFR threshold at \ssf = -10.16
yr$^{-1}$, defining the 25\% of all galaxies with a higher value of
SSFR as star-forming.  We note that our results do not qualitatively
change even if the SSFR threshold is changed to the 50-percentile
value (\ssf = -10.45 yr$^{-1}$). We thus classify the galaxies in the
4 quadrants of Fig.~\ref{ssf-gr} as red star-forming, red sequence,
blue passive and blue star-forming, anti-clockwise from top
right. Table~\ref{tab:numbers} gives the number of galaxies in our
cluster sample in each category, and also the corresponding numbers in
the whole of SDSS DR4 spectroscopic catalogue in the same redshift and
magnitude range as our cluster galaxy sample.

\begin{table}
\label{ngal}
\centering
\caption{Number of galaxies in the four categories} 
\begin{tabular}{c|c|c|c|c|}
\hline
              & red SF   & red sequence  & blue passive  & blue SF \\ \hline
 Clusters     &  539     & 3998          & 538           & 935     \\ 
 Entire SDSS  &  24039   & 163904        & 44877         & 113909  \\ \hline
\end{tabular}
\label{tab:numbers}
\end{table}

\subsection{Spatial distribution of galaxies in clusters}
\label{sec:radial}
        
 \begin{figure}
 \centering{
 {\rotatebox{270}{\epsfig{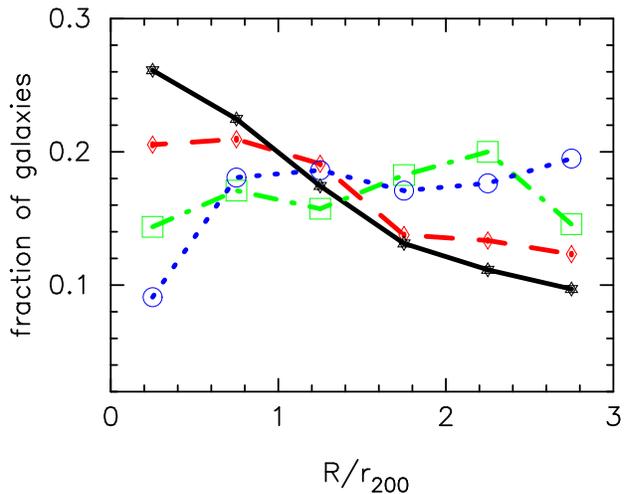}}}}
\caption{The radial distribution of galaxies in the four categories
  defined in \S\ref{sec:categories} is plotted as a function of
  cluster-centric radius, scaled by $r_{200}$.  Each distribution is
  normalized to unity.  The Poisson error on each point is smaller
  than the size of symbols. The fraction of red sequence galaxies
  {\it{(black)}}, as expected, increases toward the cluster centre,
  and the red star-forming galaxies {\it{(red)}} follow them
  closely. The blue star-forming galaxies {\it{(blue)}} show a decline
  within $r_{200}$, while their passive counterparts {\it{(green)}}
  show a constant fraction out to $3\,r_{200}$, except for a mild
  increase $\sim\! 2r\,_{200}$. }
 \label{dist}
 \end{figure}

Having divided the galaxies into four classes, we consider their
radial distribution within the clusters to examine the possible influence
of local environment. 
Fig.~\ref{dist} shows the distribution of galaxies in all four categories
as a function of distance from the corresponding cluster, scaled
by $r_{200}$. For each category, the curve has been normalized by the
total number of galaxies in that category.

As expected, the fraction of the passive red sequence galaxies
increase towards the cluster centre. The red star-forming galaxies
closely follow the distribution, but their fraction inside r$_{200}$
does not increase as steeply as that of the red sequence galaxies,
suggesting a different evolutionary path. Here it would have been very
interesting to examine the detailed difference between these two
distributions as a function of halo mass, since in both clusters and groups
the morphology and star formation properties seem to vary between
 dwarfs and giants \citep{habib04,miles06}, and/or central and satellite
 galaxies \citep{ann08}, indicating different evolutionary histories.
 But in this work, over the redshift range of our sample, we have to be
 confined mostly to $M_\ast$ and brighter galaxies ($M_r\!\le\! -20.5$).
 
At the other extreme, the blue star-forming galaxies show a constant
distribution beyond r$_{200}$, dropping steeply in the
cluster core. This is the consequence of the observed lack of neutral gas
in galaxies in the cores of rich clusters \citep[e.g.][]{gh85,coma97}.

However, the blue passive galaxies, which also have a more or less
constant radial distribution over the entire range, do not show this
rapid drop in the cores. This is consistent with the observation,
discussed below in some detail, of nuclear star formation in a
substantial fraction of brightest cluster galaxies
\citep{bildfell,reichard}, which will cause the innermost bin to be
significantly higher for this distribution than that for the blue
star-forming galaxies. It is interesting to note that the fraction of
blue passive galaxies also show a minor increment around
2r$_{200}$. Other studies have shown that in clusters, particularly
those that are fed by filaments, there is an enhancement of nuclear
star formation in the infalling galaxies beyond r$_{200}$
\citep[e.g.][]{p2,p08,mrp09}.

\section{Do the Galaxy properties derived from photometry
and spectroscopy agree with each other?}
\label{results}

One of the objectives in this paper will be to examine the hypothesis that
the galaxy populations can indeed be divided, on the basis
of a single parameter, into two classes: passive
and star-forming, and that the incidence of red star-forming or blue
passive galaxies is merely due to the scatter in the measured parameters
within reasonable errors.  As we show below, this hypothesis fails to
hold.

\subsection{The effect of extinction and abundance}
\label{analysis1}

Star-forming regions in a galaxy are often surrounded by dust, which
can significantly affect quantitative analyses based on optical
light. Conversely, the presence of dust in a galaxy would imply the
presence of star-forming regions therein.  In the absence of infrared
observations, which are often used to quantify the effect of dust in
star-forming systems \citep[e.g.][]{cal97}, optical photometry can also be
used to constrain the internal extinction in a galaxy (also see
Fig.~\ref{o-i-r}). 

In one such attempt, \citet{kauff03b} have estimated the degree of
dust attenuation in the SDSS $z$-band, by comparing model colours to
the measured colours to estimate the reddening for each galaxy. The
shape of the attenuation curve from observations is found to resemble
a power law with slope $\sim\! -0.7$ over a wavelength range from
1250-8000 \AA\ \citep{cal94}. A standard attenuation curve, of the
form $\tau_\lambda \propto \lambda^{-0.7}$, is then extrapolated to
obtain the extinction in the SDSS $z$-band. We use the $A_z$ values
from \citet{kauff03b}, expressed in magnitudes, in this section.  The
typical 1$\sigma$ error on the estimated $A_z$ is $\sim\! 0.12$ mag.
They also note that comparison of $g\! -\! r$ and $r\! -\! i$ colours
yield similar extinction values for most galaxies if the colours are
corrected for nebular emission.

\begin{figure}
 \centering{
 {\rotatebox{270}{\epsfig{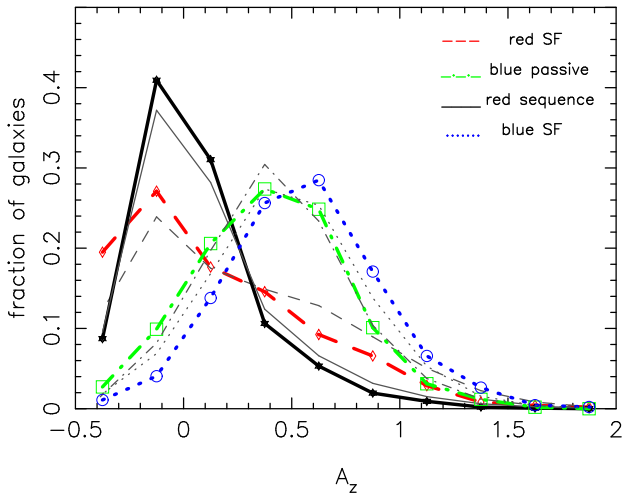}}}}
\caption{The distribution of internal extinction among our cluster
  galaxies in the four quadrants as specified in Fig.~\ref{ssf-gr}.
  Each histogram is normalized to unity.  It is clear that in general
  more `blue' galaxies are prone to internal reddening than the
  `red' ones, with star-forming ({\it{blue dotted}}) and passive
  blue galaxies ({\it{green dot-dashed}}) showing slight
  offsets. Interestingly, while $<45$\% of the red sequence galaxies
  ({\it{black solid}}) show signs of some extinction, the red
  star-forming galaxies ({\it{red dashed}}) have a bimodal
  distribution in extinction values (see text). The {\it{grey lines}}
  represent the same distributions as above, but for $>300,000$
  galaxies found in the entire SDSS DR4, divided in the four quadrants
  in the same way as described for the sample of cluster galaxies
  studied here.  Here, negative values of $A_z$ indicate the absence
  of extinction.  }
 \label{dust}
 \end{figure}

The reddening of a galaxy is the result of a combination of several
factors, involving the presence of dust and/or metals in the
interstellar medium and of old stars.  In our attempt to interpret the
apparent evidence of star formation in the fibre spectra of some red
galaxies (`red SF'), and the low values of SFR in the spectra of
some blue galaxies (`blue passive'), we plot, in Fig.~\ref{dust},
the distribution of internal extinction $A_z$ of the galaxies in the
four populations as classified in Fig.~\ref{ssf-gr}.

As expected, the `red sequence' galaxies, most of which are
passively evolving massive galaxies in the cores of clusters, show
very little extinction (negative values of $A_z$ being interpreted as
zero here). However, even among passive galaxies, there is a small
fraction that constitutes a tail extending to non-negligible
extinction values. The blue star-forming galaxies, on the other hand,
show a clear sign of attenuation going upto $1.9$~mag in the $z$-band,
with most galaxies having extinction values around $0.6$~mag for the
blue star-forming galaxies and $0.35$~mag for the blue passive
galaxies respectively.

In contrast, the red star-forming galaxies show a mix of at least two
different populations: about half of them are unattenuated (similar to
the red sequence galaxies), and the other half contributing to the
skewed high extinction end of the distribution. We employ the KMM
algorithm \citep{kmm} to quantify the existence of bimodality in
 our data.  The 
%%% add explanation of KMM algorithm %%%%%%%%%%%%%%%%%%%%%%%%
 KMM algorithm fits a user specified number 
of Gaussian distributions to the dataset,
 calculates the maximum likelihood estimate of their means and
 variances, and assesses the improvement of the fit over that provided
 by a single Gaussian.
 The KMM likelihood ratio test statistics (LRTS) used here is a
 measure of improvement in using two Gaussian distributions (since we
 are testing bimodality) over a 1-mode fit.

 For the distribution of extinction, in the red star-forming galaxies 
 the KMM test gives the value of likelihood ratio (LRTS) to be 75,
when using a bimodal fit to the data than using a unimodal fit, with
nil probability of the null hypothesis being satisfied. The 2-modes of
the internal extinction distribution are found to be centred around
A$_{z}$=-0.06 (read zero extinction) and 0.72 mag respectively for the
sample of cluster galaxies. Interestingly, we find that even by
changing the selection criteria for defining the red star-forming
galaxies to a higher value of log SFR/M$^*$ = -10 yr$^{-1}$, the KMM
test still yields statistically significant result in favour of a
bimodal fit to the data (LRTS=21 at a significance level $5\times 10^{-5}$).

Along with the distribution of extinction values for the cluster
galaxies, we also show the corresponding distributions for $>$300,000
galaxies, with the same magnitude and redshift range as our cluster
sample, drawn from the entire SDSS DR4 spectroscopic catalogue (grey
thin lines).  For this much larger sample, the histograms of
extinction values are not significantly different from those for the
cluster sample, except for the red star-forming galaxies, where
cluster galaxies seem to have relatively fewer galaxies with high
extinction values than similar galaxies in the larger all-SDSS sample.

 \begin{figure}
 \centering{
 {\rotatebox{270}{\epsfig{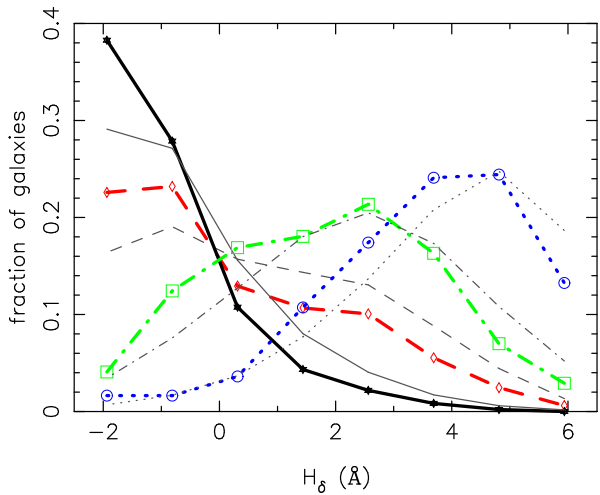}}}}
\caption{The distribution of the (emission corrected) H$_\delta$
  equivalent width (EW) for galaxies in our cluster sample, with line
  styles the same as in Fig.~\ref{dust}. As expected, the blue
  star-forming galaxies have the highest absorption, showing the presence
  of $<1$ Gyr old stellar population/s, while the red sequence
  galaxies show little evidence for absorption in H$_\delta$. The blue
  passive galaxies have a broad distribution with equal probability of
  finding H$_\delta$ in emission (negative EW) or absorption. The
  distribution for red star-forming galaxies is bimodal, which, in
  conjunction with the distribution of their internal extinction
  (Fig.~\ref{dust}), indicates the presence of two distinct galaxy
  populations.}
 \label{hd}
 \end{figure}
 
 \begin{figure}
 \centering{
 {\rotatebox{270}{\epsfig{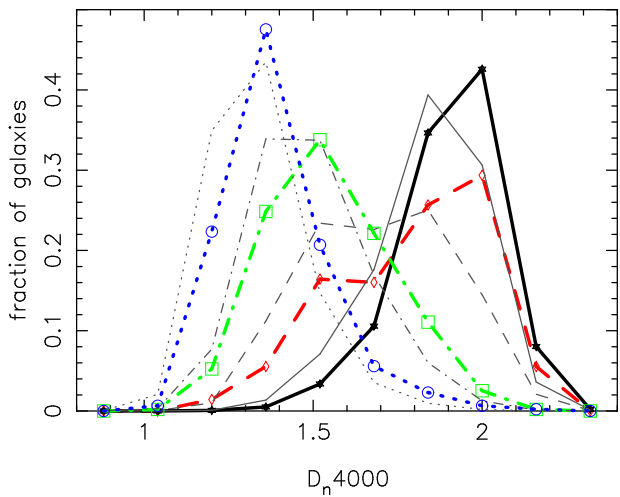}}}}
\caption{The distribution of the (NeIII continuum corrected) D$_n$4000 values,
  with line styles the same as in Fig.~\ref{dust}. As expected, the red
  sequence galaxies
  have the largest values, while the blue star-forming galaxies the
  smallest values for $D_n$4000 respectively.  The blue passive
  galaxies have a small offset ($\sim 0.2$) from their star-forming
  counterparts, revealing the presence of slightly older stellar
  populations. The histogram for red star-forming galaxies has a
  statistically significant bimodality (see text), similar to that
  seen in Figs.~\ref{dust} \& \ref{hd}.}
 \label{d4000}
 \end{figure}

\subsection{H$_\delta$ and the 4000\AA\  break}
\label{analysis2}
 
In order to probe the presence of young ($<1$ Gyr old) stellar
populations among galaxies with an unusual combination of SFR/M$^*$
and colour, we plot the distributions of the EW of the H$_\delta$
absorption line, and of the D$_n$4000 values in Figs.~\ref{hd} and
\ref{d4000} respectively. The D$_n$4000 is the strongest discontinuity
occurring in the optical spectrum of a galaxy, mainly due to the
presence of ionised metals. On the other hand, H$_\delta$ absorption
is detected after most of the massive hot stars
 have finished evolving on the main sequence, \ie\ at least 0.1-1 Gyr after a
starburst is truncated.  Thus the H$_\delta$ EW is a measure of the
age of the youngest stellar population in a galaxy, and has been
extensively used to estimate the mean stellar ages as well
\citep{wo}. Even though several studies have shown that the Balmer
emission due to HII regions, AGN and/or planetary nebulae can fill in
the underlying Balmer absorption lines, causing age estimates to be
spuriously high \citep{trager00,prochaska07}, the
H$_\delta$ EW remains a popular stellar age indicator because it is
less affected than the lower order Balmer lines, due to the steep
Balmer decrement in emission \citep{osterbrock89,wo}.

Here we use the measured values of the (emission corrected) H$_\delta$
EW, and the (NeIII continuum corrected) D$_n$4000 from
\citet{kauff03b}, who, for the latter, have adopted a modified version
of the original definition \citep{bruzual83} of the D$_n$4000
introduced by \citet{balogh99}. A stronger D$_n$4000 is an indicator
of a metal-rich inter-stellar medium, which, together with the
H$_\delta$ EW, indicates whether the galaxy has been forming stars
continuously (intermediate/high D$_n$4000 plus low H$_\delta$ EW) or
in bursts (low D$_n$4000 plus high H$_\delta$ EW) over the past 1-2
Gyr \citep{kauff03b}. We note that the observational error in
D$_n$4000 index is very small, typically $\sim$0.02, but the same is
not true for the H$_\delta$ EW, which has an uncertainty of 1-2 \AA~in
measurement \citep{kauff03b}. \citet{kauff03b} discuss the physical
constraints leading to these uncertainties, and show that the
H$_\delta$ index can assume only a selected range of values for all
possible star formation histories. The large observational errors on
the H$_\delta$ EW imply that the estimated value of this index is
constrained by the parameter space spanned by the acceptable models.
  
 The red sequence galaxies in our sample show H$_\delta$ in emission, and
 yield large values for the D$_n$4000, confirming the fact that their
 main ingredient is a population of old, passively evolving stars,
 which determines the overall galaxy colours and dominates the
 spectra. The high observational uncertainty in the EW(H$_\delta$)
 implies that it is hard to determine whether H$_\delta$
 is actually occurring in emission, or the negative EWs are a result of a high
 correction factor in individual galaxies. This is a 
 common problem with the use of H$_\delta$ index in individual systems,
 but this might not be an important
 issue for this statistical study. 
 The blue star-forming galaxies, in sharp contrast, yield
 small values for the D$_n$4000 and strong H$_\delta$ in
 absorption. The blue passive galaxies follow the blue star-forming
 galaxies, with the mean value for the D$_n$4000 systematically higher
 by $\sim0.2$, but with a very broad distribution of H$_\delta$ EW,
 revealing a mix of galaxies with a large scatter in the ages of their
 youngest stellar populations. In red star-forming galaxies, the
 distributions of H$_\delta$ EW as well as the D$_n$4000 are
 bimodal. The KMM test prefers a 2-mode fit over a 1-mode fit with
 likelihood ratios of 78 and 84, for H$_\delta$ EW and D$_n$4000
 respectively, with zero probability for the null hypothesis
 in both cases.
 
 \begin{figure}
 \centering{
 {\rotatebox{270}{\epsfig{file=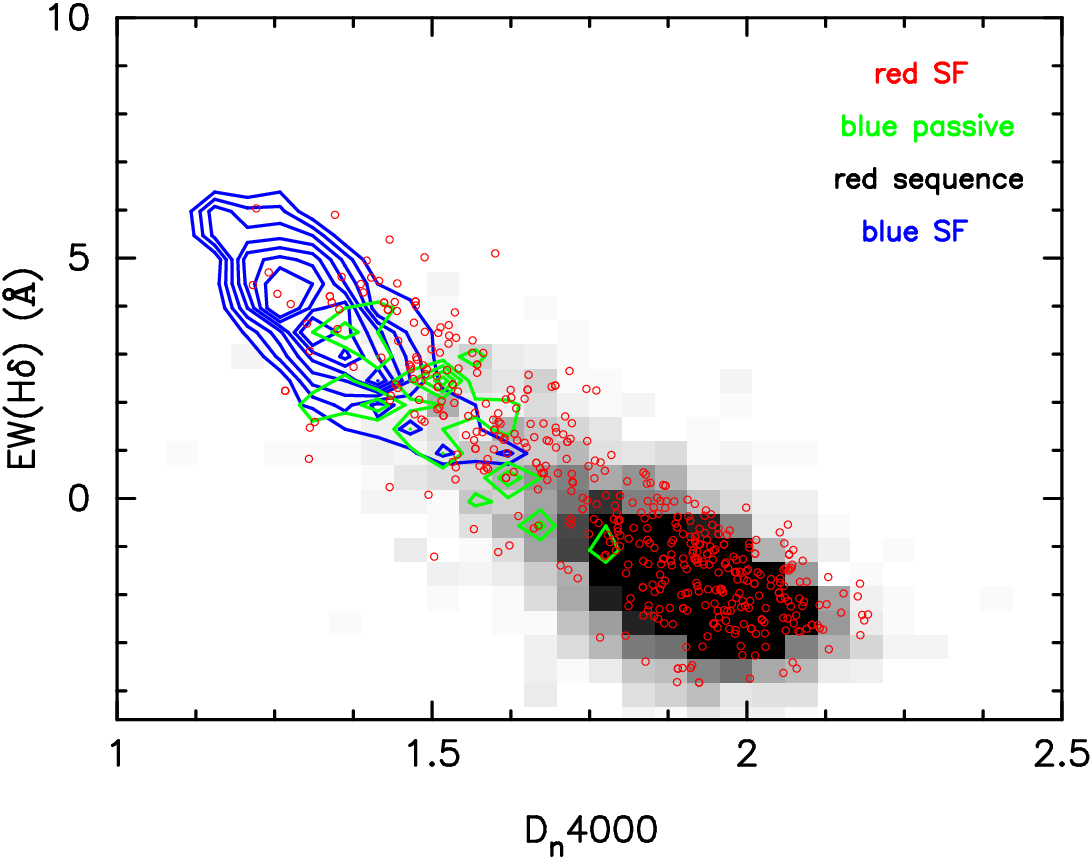,width=6.5cm}}}}
\caption{
  The relation between the equivalent width of the $H_\delta$ line
  EW(H$_\delta$) and the 4000\AA\ break index D$_n$4000, for all
  galaxies in our sample.  These two indices together form a robust
  diagnostic of the star formation history of a galaxy, irrespective
  of the stellar ages and metallicity \citep[][also see
  text]{kauff03a}.  The red sequence galaxies {\it{(grey scale)}} form
  a core in the passive region of the plane, while the blue
  star-forming galaxies {\it{(blue contours)}} occupy the region
  indicating recent and ongoing star formation.  The blue passive
  galaxies {\it{(green contours)}} closely follow the blue star
  forming galaxies, but with somewhat higher D$_n$4000 and low
  EW(H$_\delta$), indicating recent truncation of star formation. The
  innermost contours represent 24 galaxies in both cases, and the
  number decreases by 3 at every successive level of contour.  The red
  star-forming galaxies seem to span the entire range in both indices
  defining this space, a relatively large fraction of them being
  passively evolving just like the red sequence galaxies, but a
  non-negligible fraction intruding the space dominated by the blue
  star-forming galaxies. }
 \label{hd-d4}
 \end{figure}

 We also check whether this bimodality is a result of the threshold
 chosen to define whether a galaxy is star-forming or not.  As for the
 internal extinction, the KMM test results remain statistically
 significant, for both H$_\delta$ EW and the D$_n$4000, when the red
 star-forming galaxies are selected with an even higher SSFR threshold
 (log SFR/M$^*$ = --10 yr$^{-1}$), showing that these results are not
 sensitive to how the four categories are chosen, within reasonable
 limits. The above results become even more convincing when we plot
 the four different galaxy populations on the plane defined by the two
 indices D$_n$4000 and EW(H$_\delta$)
 (Fig.~\ref{hd-d4}). \citet{kauff03a} have shown that these two
 indices taken together can very effectively determine the star
 formation history of a galaxy. As shown in Fig.~\ref{hd-d4}, this
 indeed seem to be the case.  The blue star-forming and the passive,
 red sequence galaxies occupy the opposite ends of the space. But
 whilst the blue passive galaxies seem to follow the blue star-forming
 galaxies, although with somewhat lower EW(H$_\delta$) and higher
 D$_n$4000, the red star-forming galaxies span the entire range of
 values in both indices, as would be expected given their bimodal
 distribution.

 \begin{figure}
 \centering{
 {\rotatebox{270}{\epsfig{file=figure-8.ps,width=6.5cm}}}}
\caption{The cumulative distributions of $g\!-\! r$ aperture colour, 
   measured within the
  3$^{\prime\prime}$ SDSS fibre, relative to the overall colour of the
  entire galaxy.  At $z \!\sim\! 0.1$, $>60$\% of the light in the
  fibre comes from the bulge and hence the fibre colour can be assumed
  to be a proxy for the colour of the bulge. The slope of the
  distributions for blue galaxies is much steeper than that of the red
  ones, showing the large spread in the relative colour, irrespective
  of the SSFR. Although the red galaxies have a relatively small
  spread in colour, a significant number of red star-forming galaxies
  have bluer bulges. This could be the signature of the presence of
  active nuclei supporting star-formation activity in the core of the
  galaxies, while the overall galaxy colour remains dominated by the
  older stellar population elsewhere in the galaxy. }
 \label{col-ratio}
 \end{figure}

 \begin{figure*}
 \centering{
 {\rotatebox{0}{\epsfig{file=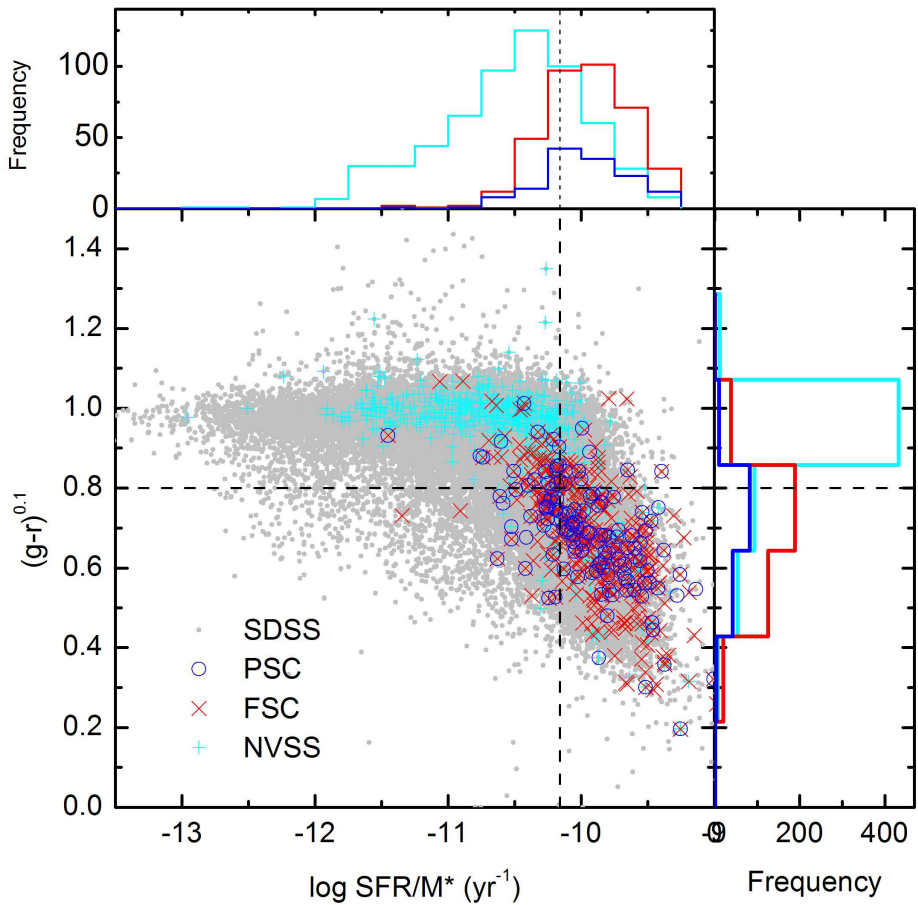,width=18cm}}}
 \caption{The central panel of this plot shows the $(g\!-\! r)^{0.1}$
   colour of all the spectroscopically covered SDSS DR4 galaxies ({\it{grey dots}})
   in the colour-SSFR space (same as Fig.~\ref{ssf-gr}), found
   in the same redshift range and magnitude limit as our cluster galaxy
   sample.
   Over-plotted are 1.4~GHz radio sources from the NVSS ({\it{cyan}}),
   {\it{IRAS}} point sources (PSC, {\it{blue}}) and IRAS faint sources (FSC,
   {\it{red}}) respectively. The top and the right panel show the
   histograms of these quantities along the two axes. As expected,
   most of the radio-loud AGN concentrated along the red sequence, and
   IR bright sources in the blue star-forming galaxies. It is
   interesting to note that the red star-forming galaxies (top right
   quadrant, central panel), a large fraction of which are expected to
   be dust-reddened, do not seem to have many IRAS sources. This
   supports our conclusion here that a non-negligible fraction
   of the red star-forming galaxies obtain their colour from the
   presence of excessive metals.
 \label{o-i-r}}}
 \end{figure*}

\subsection{AGN and star formation} 

Figs.~\ref{dust}, \ref{hd}, \ref{d4000} and \ref{hd-d4} clearly show that while
most of the cluster galaxies follow the classical picture, blue
colours correlating with higher SSFR, and redder colours with
negligible star formation activity, there is a non-negligible fraction
of galaxies in either class which deviate from the norm.  It is thus
crucial to understand the properties of these galaxies and identify
the cause of their occurrence. In order to do so, we make use of the
classification of galaxies, based on the BPT diagram, as star-forming,
composite, AGN and those without emission lines (unclassified), as
mentioned above in \S\ref{galaxy-data}.

Ignoring, for the time being, the unclassified galaxies for which the
nature of star formation activity cannot be conclusively determined
from the spectroscopic data, we find that $>$30\% of the red
star-forming galaxies and $\sim$50\% of the red sequence galaxies are
classified as AGN. This agrees with the well-known observation that
AGN are predominantly hosted by the massive elliptical galaxies. In
Fig.~\ref{o-i-r}, we plot 1.4GHz radio sources \citep[from
NVSS,][]{best} and IRAS 60$\mu$m sources found in the SDSS DR2
\citep{cao}, along with all the galaxies found in the same redshift
range ($z\!\leq\! 0.12$) and magnitude limit (M$_{r}\!\leq\! -20.5$)
in our galaxy sample. The radio-loud AGN are clearly seen to be hosted
mostly in the red sequence galaxies. Interestingly, not many of the
red star-forming galaxies are found in the IRAS $60\mu$m source
catalogues, disagreeing with the usual expectation that the colour
of the red star-forming galaxies is a result of the presence of dust
around star-forming gas clouds. We further explore this issue in
\S\ref{rs-sec}.

\subsection{Aperture vs global colours} 
 
An SDSS spectrum comes from a fixed aperture, which is equal to the
diameter of the fibres (3$^{\prime\prime}$) used in the
spectrograph. Since this aperture would encompass a varying fraction
of the light of a galaxy depending upon its size and distance, going
up to $\sim$40-60\% at $z\!=\! 0.1$, the spectrum might not, in most
cases, be a true representation of the galaxy's mean overall star
formation activity.  In this work, we compare photometric colours
and spectroscopically derived specific star formation rates, which
come from independent observations of the same galaxies. It would be
interesting to see how the colour derived from the fibre aperture, as
opposed to the value averaged over the entire galaxy, relates to SSFR.
For instance, if an early-type galaxy has significant star formation
in the core, we would expect to see a bluer core, while the overall
photometric colour would be dominated by the older stellar populations
elsewhere in the galaxy.

In Fig~\ref{col-ratio}, we plot the difference between the colour
measured within the fibre and the overall colour of the galaxy,
normalised by the overall colour, separately for galaxies belonging to
the four categories. A noticeable characteristic of this plot is that
irrespective of their star formation properties, the blue and red
galaxies have very different slopes for this cumulative
distribution. The steeper slope for the blue galaxies is a consequence
of their wider range in colour. While most of the galaxies have fibre
colours that are redder than the overall galaxy colours, $\sim$15\% of
the blue star-forming and $\sim$15\% of the red star-forming and red
sequence galaxies have bluer cores. This suggests the presence of
centrally concentrated star formation in these galaxies. It must be
noted that although relative fibre colour is a good first
approximation to the colour of the bulge of a galaxy, in this work
where we stack galaxies over a range of redshift, the ratio of central
and overall colour is not derived on the same physical scale for all
galaxies. This result should therefore be considered to be a
qualitative one.

\section{Discussion}
\label{discussion}

Numerous studies of the evolution of galaxies have made use of either
global photometric properties such as colour \citep[\eg][]{bo},
morphology \citep{d2} or luminosity of a galaxy, or spectroscopic
properties such as circular velocity \citep[\eg][]{graves}, the
D$_n$4000, EW of H$_\delta$ absorption line \citep[\eg][]{balogh99},
 emission lines of H$_\alpha$ or OII
\citep[\eg][]{lewis,g3,b7,h1} or integrated SFR and/or SSFR (B04). Often samples
are chosen based on colour or star formation properties.  Are the
properties derived from photometry and from spectra correlated enough
to yield similar results?

Blue galaxies are thought to be of late-type morphology and and have
significant star formation activity, while red galaxies are early-type
and show little or no evidence of current star formation.  However,
with the recent availability of extensive multi-wavelength photometric
databases, and of large optical spectroscopic surveys, several authors
\citep[\eg][]{wolf,bildfell,haines08,koyama,saintonge,gfs,wolf09} have 
suggested that optical colour may not necessarily be a good
representation of a galaxy's evolutionary state. In this paper, we
have explored both photometric colours and spectroscopic specific
star formation rate from optical
observations of galaxies, in a wide range of environments, in and near
rich clusters in the local Universe, to examine cases where the usual
correlation between the two fails to hold.
 
We find that $>$80\% of the red galaxies ($(g\!-\! r)^{0.1}\geq 0.8$)
that are classified to be part of the red sequence of clusters, on the
basis of photometric colours, also have very little or no star
formation (\ssf$\leq$-10.16 yr$^{-1}$, see Fig.~\ref{ssf-gr}).
Similarly, 65\% of the blue galaxies found within $3\,r_{200}$ of rich
clusters also have high SSFR. However, in addition to these `normal' galaxies in
which the colour and spectroscopic SFR/M$^*$ correspond well with each other, 
a significant fraction of galaxies deviate from the norm,
and would be classified differently, if treated solely on the basis of
one of their spectroscopic and photometric properties.
 
 \subsection{Blue passive galaxies}
\label{bpsec}

The blue passive galaxies in our cluster sample (Fig.~\ref{ssf-gr})
have spectroscopic properties quite similar to their star-forming
counterparts. The colour of a galaxy cannot be expected to, and indeed
does not, correlate in the same way with environmental properties, as
the star formation properties do \citep[e.g.,][]{kauff03b}.
Our analysis shows that the values for the D$_n$4000
(Fig.~\ref{d4000}) and internal extinction (A$_z$; Fig.~\ref{dust})
for blue passive galaxies are similar to those of blue star-forming
galaxies, but the H$_\delta$ EW distribution (Fig.~\ref{hd}) is very
different for the two populations. Since our sample is chosen to be in
or near rich clusters, the discrepancy between their mean photometric
colour, and their SSFR derived from spectra, suggests two possible
scenarios: (i) these are galaxies in which star formation has been
recently ($\sim$1-2 Gyr) shut off due to the impact of local and/or
global environment \citep[\eg][]{p1,p2,p08,mrp09}, 
or (ii) these are
bulge dominated galaxies where the mean colour is dominated by the
young stellar populations distributed in the disc but the spectrum
sourced from an almost dead centre shows little sign of star formation
activity. 

A galaxy that has experienced a major burst of star formation 1-2 Gyr
ago will show significant absorption in H$_\delta$, which is found in
a non-negligible fraction of the blue passive galaxies here
(Figs.~\ref{hd} and \ref{hd-d4}). This is because the A stars which
significantly contribute to the H$_\delta$ absorption, have a lifetime
of $\sim$1.5 Gyr, but the burst of star formation produces enough
ionizing radiation to fill in the H$_\delta$ absorption line. But the
absorption line shows up in the galaxy spectrum once the starburst has
been truncated/ended.

 \begin{figure}
 \centering{
 {\rotatebox{270}{\epsfig{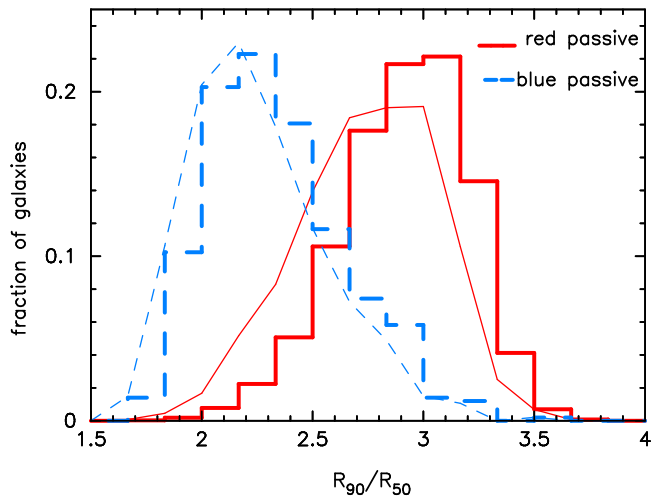}}}}
\caption{The distribution of the concentration parameter, C $\equiv$
  R$_{90}$/R$_{50}$ for the red sequence ({\it{red solid}}) and blue
  passive ({\it{blue dashed}}) galaxies respectively. The thick and
  thin lines represent the distributions with and without unclassified
  galaxies in each case respectively. The blue passive galaxies
  clearly have late-type morphologies along with bluer colours. The
  fact that these galaxies have very low SSFR thus make them
  candidates for progenitors of the red sequence galaxies, in which
  star formation has been recently ($<1$ Gyr) truncated, eventually
  leading to red colour and early-type spheroidal structure typical of
  the present day cluster red-sequence galaxies.}
 \label{bp}
 \end{figure}

\begin{table}
\label{ks}
\centering
\caption{K-S test results for distributions in Fig.~\ref{bp} \& \ref{rs} respectively} 
\begin{tabular}{l|c|c|}
\hline
 Distributions                             & D     & p      \\ \hline
 Red \& blue star-forming galaxies         & 0.67  & 0.0000 \\
 (including unclassified galaxies)         &         &        \\
 Red \& blue star-forming galaxies         & 0.60  & 0.0000 \\
 (excluding unclassified galaxies)         &         &        \\
 Red star-forming galaxies                 & 0.08  & 0.1870 \\
 (including \& excluding unclassified galaxies) &    &        \\
 Blue star-forming galaxies                & 0.02  & 0.9999 \\
 (including \& excluding unclassified galaxies) &    &        \\\hline
 Red \& blue passive galaxies              & 0.35  & 0.0000 \\
 (including unclassified galaxies)         &         &        \\
 Red \& blue passive galaxies              & 0.29  & 0.0000 \\
 (excluding unclassified galaxies)         &         &        \\
 Red passive galaxies                      & 0.08  & 0.1772 \\
 (including \& excluding unclassified galaxies) &    &        \\
 Blue passive galaxies                     & 0.01  & 0.9999 \\
 (including \& excluding unclassified galaxies) &    &        \\\hline
\end{tabular}
\end{table}

Fig.~\ref{bp} provides statistically significant evidence in favour of
(ii). As a morphological index, here we use the
concentration parameter $C\equiv R_{90}/R_{50}$, where R$_x$ is the
Petrosian radius encompassing x\% of the galaxy's light in the
$r$-band from SDSS images.  In Fig.~\ref{bp}, we plot the distribution
of $C$ for the red and blue passive galaxies respectively. The blue
passive galaxies clearly show late-type morphologies
\citep[$C\! <\! 2.6$;][]{strateva}, while their red counterparts mostly have
values of concentration typical of spheroidals. The galaxies in
the former category are similar to the passive blue spirals that have been
 studied by other authors \citep[\eg][]{goto03,wild}. 

By analysing the spatially resolved long-slit spectroscopy of a few
passive spiral galaxies selected from the SDSS, \citet{ishigaki07} find
that the H$_\delta$ absorption lines are more prominent in the outer
regions of the disc, while the nuclear regions show strongest values
for D$_n$4000. Their findings are consistent with the idea that the
star formation in passive late-type galaxies ceased a few Gyr ago.
Hence, the blue passive galaxies can be considered to be the
progenitors of their red counterparts, in which star formation has
been recently shut-off: these galaxies will eventually acquire redder
colours and early-type morphologies, over a period that depends upon
their environment \citep{goto03c}. \citet{kannappan09} have recently
identified, characterized and studied the evolution of galaxies which,
though morphologically classified as E/S0s, lie on the blue sequence
in the colour-stellar mass space, and suggest that these systems
almost always have a bluer outer disk, and that they are a population
in transition, potentially evolving on to the red sequence, or have
(re)formed a disk and are about to fall back in the class of
late-types.

 \subsection{Red star-forming galaxies}
 \label{rs-sec}

 Even more interesting are red galaxies with signs of active star
 formation in their fibre spectra. We find that 53.4\% of the 539
 galaxies in this class have significant emission lines on the basis
 of which their SSFR has been evaluated.  These red star-forming
 galaxies are made up of at least two kinds of galaxies 
 (chosen from simple limits in colour and SSFR).  One kind appears to be
 very similar in photometric properties to the red sequence galaxies
 but with spectroscopically derived properties consistent with their
 late-type counterparts, while the other class has both photometric
 {\it{and}} spectroscopic properties similar to the red sequence
 galaxies, yet has a high SSFR.

 The origin of the latter class is not clear. It is likely that these
 galaxies are currently experiencing a starburst, which has 
 started very recently ($\lesssim$0.5 Gyr). Stellar populations
 resulting from such a starburst would be detected in spectral indices
 such as H$_\alpha$ EW (leading to high values of SFR/M$^*$), but will
 not dominate the D$_n$4000 and the H$_\delta$ EW for another
 $\sim$0.5--1~Gyr.  The former sub-population is most likely to be
 bright cluster galaxies (BCGs), with nuclear star-formation linked
 with the AGN \citep[see][]{goto06,bildfell,reichard}. In the following we
 turn our attention to this sub-population.

 \subsubsection{Star formation at the core of giant ellipticals}

 The hypothesis that AGN and star formation activity in the core of a
 galaxy may be linked is supported by the fact that $\sim$30\% of the
 red star-forming galaxies with emission lines are classified as AGN
 on the BPT diagram. The fact that 50\% of the emission-line red sequence
 galaxies (defined according to Fig.~\ref{ssf-gr})
 are also classified as AGN does not contradict this scenario because
 either (i) the red sequence AGN galaxies in general have high stellar
 mass, and so can have low SSFR even though they have significantly
 high SFR, or (ii) they do not have enough cold gas to form stars,
 because of the presence of an active nuclei. 

 A significant fraction of giant ellipticals show evidence of ongoing
 star formation or signs of recent ($<$2~Gyr) star formation in their
 cores, particularly in environments where the likelihood of recent
 mergers is high \citep[e.g.,][]{mcd06,nolan06,nolan07}. For a sample of
 BCGs in 48 X-ray luminous clusters, \citet{bildfell} find that 25\%
 of the BCGs have colour profiles $(g\! -\! r)$ that turn bluer toward
 the centre. Our radial distribution of galaxies in this category 
(Fig.~\ref{dist}), compared to the distribution of blue 
star-forming galaxies, is consistent with this (see \S\ref{sec:radial}).

Elsewhere, \citet{gallazzi09} analyse an extensive
 dataset, covering UV to IR SEDs for galaxies in the Abell 901/902
 cluster pair, to show that $\sim$40\% of the star-forming galaxies
 residing in intermediate to high density environments have optically
 red colours. They show that these galaxies are not starbursts, and
 have SFRs similar to or lower than that of their blue counterparts.
 Whether the occurrence of blue cores in red sequence galaxies is
 linked to their nuclear activity, and whether the nuclear activity
 itself is modulated by the immediate environment of a galaxy (such as
 the presence of close neighbours), remains a debatable issue. We
 intend to return to this in future work involving a more detailed
 study of nuclear starbursts (also see \citealp{li,reichard}).

The scenario discussed in (ii) above complements the blue passive
 galaxy population discussed in \S\ref{bpsec}-- massive galaxies
 without much cold gas could be formed in dry major mergers of the
 passive blue progenitors, while high speed encounters amongst them in
 moderately dense environments, such as in the infall regions of clusters
 \citep[\eg][]{p08,mrp09}, could lead to excessive
 loss of cold gas, leaving the galaxy to evolve passively thereafter.

 In a spatially resolved spectroscopic study of 48 early-type
 galaxies, the SAURON team \citep{sauron} simultaneously observed the
 CO(1-0) and CO(2-1) lines with 28\% detection rate (12/43)
 \citep{combes07}. In this sample they find that the CO-rich galaxies
 show strongest evidence of star formation in the central ($\sim$4x4
 arcsec$^2$) regions and also tend to have higher H$_\beta$ and lower
 Fe5015 and Mg$b$ indices. The results presented in this work also
 support various studies based on the UV data obtained by the Galaxy
 Evolution Explorer ({\it{GALEX}}) which suggests that a significant fraction
 ($\sim$15\%) of bright ($M_{r}\!<\! -22$) early-type galaxies show
 signature of recent ($\lesssim$1 Gyr old) star formation
 (\eg\citet{yi05}; also see \citet{crocker09,gallazzi09}).

 \begin{figure}
 \centering{
 {\rotatebox{270}{\epsfig{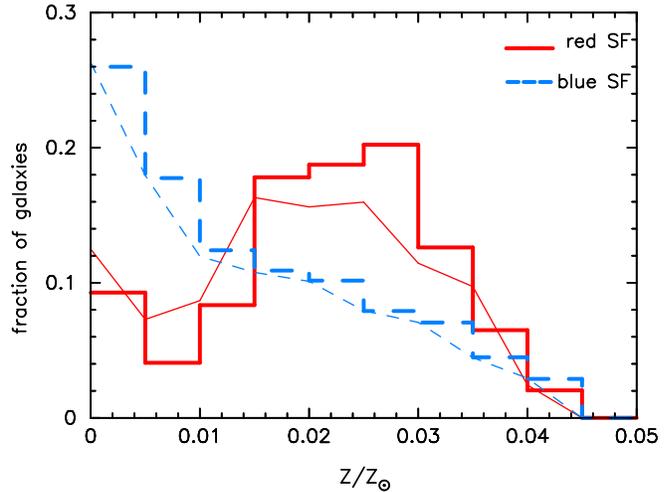}}}}
\caption{The distribution of stellar metallicities
  \citep[from][]{gallazzi} for the red star-forming ({\it{red solid}}) and
  blue star-forming ({\it{blue dashed}}) cluster galaxies in our sample. The
  thick and thin curves are distributions including and excluding the
  unclassified galaxies respectively in each case. The distributions
  evidently show that the stars in the red star-forming galaxies are
  much more metal-rich with respect to their bluer counterparts. Given
  that the internal extinction in galaxies belonging to the two
  categories does not show a significant difference (Fig.~\ref{dust}),
  presence of metal-rich stellar populations in the red star-forming
  galaxies seems to be a likely cause of their redder colours.}
 \label{rs}
 \end{figure}

 \subsubsection{Stellar and gas phase metallicities}

 Since star formation in a galaxy is often associated with the
 presence of dust, one would naively expect these red star-forming
 galaxies to be the dusty starburst galaxies widely discussed in the
 literature
 \citep{wolf,bildfell,koyama,saintonge,gallazzi09,wolf09}. However, we
 find that only $\sim 50$\% of the galaxies in this class show
 presence of any optical extinction (Figs.~\ref{dust} \&
 \ref{o-i-r}). So why are the rest of the galaxies red?
 
 To address this question, we explore two other factors affecting
 galaxy colours, namely age and metallicity.  We use stellar
 metallicities and mean luminosity-weighted stellar ages from the work
 of \citet{gallazzi}, which are estimated from 150,000 Monte Carlo
 simulations of a full range of physically plausible star formation
 histories. These models are logarithmically distributed in a
 metallicity range of 0.2--2.5 $Z_\odot$. In a given model, all stars
 have the same metallicity, interpreted as the `luminosity-weighted
 stellar metallicity'. For each model in the library, \citet{gallazzi}
 measure the strength of the D$_n$4000, and the H$_\beta$, H$_{\delta
   A}$+H$_{\gamma A}$, [Mg$_2$Fe] and [MgFe]$^\prime$ spectral indices
 in the same way as they are measured in the SDSS spectrum of a
 galaxy. By comparing the strengths of these indices in the observed
 spectrum to the model spectra in their library, the probability
 density functions (PDFs) of several physical parameters, such as age
 and metallicity of a galaxy, are constructed.

 We use the $r$-band luminosity weighted ages (hereafter referred to
 as `mean galaxy ages') and stellar metallicities derived from these
 PDFs\footnote{http://www.mpa-garching.mpg.de/SDSS/DR4/Data/stellarmet.html}.
 For galaxies in our cluster sample, we find that the red star-forming
 galaxies are systematically older than their blue counterparts by
 5-7~Gyr. In Fig.~\ref{rs}, we compare the distributions of stellar
 metallicities of the red and blue star-forming galaxies. We confirm that
 the two distributions are statistically different by using
 the Kolmogorov-Smirnoff (K-S) statistics, which finds the probability of two datasets
 being drawn from the same parent population. The K-S test finds the
 maximum difference between the cumulative frequency distributions of the red and
 blue star-forming galaxies to be $D\!=\!0.35$, with the probability of the null
 hypothesis being satisfied to be essentially zero. 

 Further support for the presence of excessive metals in the red
 star-forming galaxies is provided by a comparison between the
 gas-phase metallicities (log[O/H]$+$12) estimated by
 \citet{tremonti}\footnote{http://www.mpa-garching.mpg.de/SDSS/DR4/Data/oh\_catalogue.html},
 who have compiled a sample of $\sim$53,000 star-forming galaxies
 from the entire SDSS DR4 (selected to have H$_{\alpha}$, H$_\beta$ \&
 [NII] emission lines at $>\! 5\sigma$ detection). We find that the
 cumulative distributions of the gas-phase metallicities of the red
 and blue populations are statistically different. The K-S test yields
 $D\!=\!  0.18$ at a significance level of 0.043. Though the two
 samples compared here have widely different numbers of galaxies, our
 conclusion is consistent with that of \citet{ellison}, who show that
 for a sample of 43,690 galaxies, selected from the SDSS DR4, the
 gas-phase metallicities of galaxies with high SSFR are lower by 0.2
 dex, compared to those with low SSFR at fixed stellar mass. The fact
 that their metallicities are estimated by a comparable but
 independent method (see \citet{kd} for details), further strengthens
 our result.
 
 Around 35\% of all red star-forming galaxies are classified as
 star-forming or composite galaxies from their spectra. This is in
 agreement with the results of \citet{gfs} who also find $\sim$30\%
 of nearby red sequence galaxies from SDSS show presence of emission
 lines in their spectrum. They analyse LINER like emission in these
 galaxies to show that they are systematically younger by 2.5-3 Gyrs
 than their quiescent counterparts without emission at fixed
 rotational velocity. At $z\!\sim\!0.1$, \citet{tremonti} have shown
 that the metallicity is well correlated with the stellar mass of the
 galaxy. These quantities correlate well for 8.5$\leq$log
 M$^{*}\leq$10.5 M$_\odot$, but the relation flattens thereafter. The
 fact that $\geq$60\% of our red galaxies have log M$^*$ in the range
 10.6-10.9 M$_\odot$ supports the hypothesis that a significant
 fraction of the red star-forming galaxies obtain their red colours
 from the residual older generations of stars.

 We have also shown that $\sim$50\% of all the red star-forming
 galaxies are likely to be attenuated by $>$0.2 mag in the SDSS
 $z$-band (Fig.~\ref{dust}), and have bimodal H$_\delta$ EW and
 D$_n$4000 distributions (Figs.~\ref{hd}, \ref{d4000} and \ref{hd-d4}) which are
 very different from that of the typical red sequence galaxies. As
 discussed in \S\ref{analysis1} and \S\ref{analysis2}, the red
 star-forming class of galaxies remain bimodal in the distributions of
 A$_z$, H$_\delta$ and D$_n$4000, even if the selection criterion
 based on SFR/M$^*$ is changed to a significantly different value
 (e.g. $\sim$15-percentile, corresponding to log SFR/M$^*$ = -10
 yr$^{-1}$). This clearly indicates that the component of red
 star-forming galaxies that are similar in properties to the red
 sequence galaxies are not merely a result of scatter from the red
 sequence (quadrant 1; Fig.~\ref{ssf-gr}).

 \subsection{Implications for the Butcher-Oemler effect}
 
 \citet{bo} found that clusters at moderate to high redshift contain
 an excess of blue galaxies, compared to their low-redshift
 counterparts.  For the redshift regime of our sample
 ($z\!\sim\!0.1$), they found a uniform blue fraction f$_b$ of
 $\sim$0.03 in all clusters within R$_{30}$, the radius containing
 30\% of the cluster's red sequence population.  This exercise has
 been repeated several times in recent years for many samples of
 clusters \citep{bo,m3,e2,d3}. Most studies of this kind define
 `blue' galaxies in terms of a broadband colour that is bluer by
 0.2~mag than that of the cluster's red sequence galaxies. From the
 above discussion, it is clear that this cut leaves out a significant
 fraction of star-forming galaxies that have redder broadband colours,
 and classifies some passively evolving galaxies as `star-forming'.

 \citet{m3} found the blue fraction of galaxies to be
 f$_b$=0.03$\pm$0.09, similar to \citet{bo}, for 44 Abell clusters at
 0.03$\leq$z$\leq$0.38 within a fixed cluster-centric distance of
 0.7~Mpc, which translates into the mean R$_{30}$ for most Abell
 clusters below $z\lesssim 0.1$. However, \citet{m1} measured the
 fraction to be $(1.24\pm 0.07)z-0.01$ for clusters at $z\!\leq\!
 0.25$, again for galaxies within a fixed cluster-centric aperture of
 0.7~Mpc. Elsewhere \citep{e2,d3}, various multiples of
 cluster-centric radius, scaled with r$_{200}$, have been used to show
 the effect of chosen aperture size on the blue fraction. \citet{e2}
 suggest that the origin of the Butcher-Oemler effect lies in the fact
 that the relative fraction of `blue' galaxies on the outskirts of the
 clusters at higher redshifts is higher than in their local
 counterparts. If confirmed, this result can have a critical
 impact on the study of evolution of galaxy properties.
 It is worth mentioning here that even though most of these studies
 use optical spectra for most or all of their sample (except for
 \citet{bo}), the spectroscopic information is used only for assigning
 cluster membership.

 Recently, with the increasing availability of large multi-wavelength
 datasets, the possibility of the Butcher-Oemler effect being observed
 in other fundamental galaxy properties
 \citep[\eg~morphology,][]{goto03b}, or in non-optical data
 \citep[\eg~mid-IR, ][]{saintonge}, is being explored. The study of
 \citet{wolf} unraveled the presence of young stars and dust in the
 red sequence galaxies in a cluster pair at z$\sim$0.17, while
 \citet{bildfell} find that the colour profile of 25\% of the
 brightest cluster galaxies (BCGs) turn bluer by 0.5--1.0 mag towards
 their centres, compared to the average colours seen in the cluster's
 red sequence. All of this has serious implications for
 Butcher-Oemler like effects based on global photometric properties.

Our sample provides some additional insight to this discussion.  The
presence of metals, and of colour gradients
related to SFR in a galaxy, will affect the estimated
fraction of `star-forming' galaxies in low redshift clusters. The fact
that nearby cluster galaxies are more metal-rich than those
in high-redshift clusters, will artificially enhance the
fraction of passive galaxies that are selected solely on the basis of
their colour. Such an effect is usually not accounted for while comparing the
blue/star-forming galaxy fraction in clusters at different
redshifts. A quantitative analysis of this effect requires a
consistent study of an unbiased sample of mutually comparable galaxy
clusters spanning a wide redshift range, for which both photometric
and spectroscopic multi-wavelength data is available. We leave this
for a later consideration.

 \section{Summary and Conclusion}
 \label{conclusions}

 We divide a sample of $>$6,000 galaxies found in or near rich Abell
 clusters (z$\leq$0.12) in the SDSS spectroscopic catalogue, into four
 populations, defined by placing simple limits in $(g-r)^{0.1}$ colour
 (derived from photometry) and specific star formation rate (derived
 from the detailed modelling of optical spectra). While most blue
 galaxies have evidence of star formation, and red galaxies do not,
 this identifies two significant populations of blue passive galaxies
 and red star-forming galaxies.

 We trace the spectroscopic and photometric properties of galaxies in these
 four sub-samples in detail, using data available from SDSS DR4. The
 main results of our analysis can be summarised as follows:

 \begin{itemize}
 \item For the blue passive galaxies, the dust content and values for
   the D$_n$4000 are similar to those of the blue star-forming
   galaxies, indicating similar metal content, mean ages and internal
   extinction.

 \item Since the blue passive galaxies have a broad distribution of
  H$_\delta$ EW, along with late-type morphologies, they appear to be
   similar to the passive blue spirals studied by
   other authors. Our results support the hypothesis that these
   galaxies could be the progenitors of the red sequence galaxies in
   which star formation has been recently shut off due to the impact
   of local/global environment.

 \item An intriguing find of this study is that only $\sim$50\% of the
   red star-forming galaxies have significant values of extinction,
   indicative of the presence of dust. 

  \item By comparing the stellar
   metallicities, gas-phase metallicities and mean ages of the red and
   blue star-forming galaxies, we find statistically significant
   evidence in support of the fact that, for at least half of the red
   star-forming galaxies, the red colours result from a relatively
   metal-rich stellar population. Our argument remains valid even if
   the criterion for classification as star-forming based on SSFR is
   changed within reasonable limits (15-50 percentile, instead
   of the 25-percentile used in this work).

 \item A significant ($\sim$15\%) fraction of the red star-forming 
   galaxies have blue cores, indicating recent star
   formation which might be linked to the AGN activity in these
   galaxies.
   
 \item The population of passive red sequence galaxies increase toward the cluster
 centre, closely followed by the red star-forming galaxies, except that the latter
 does not rise steeply in the cluster core.
 
 \item The blue star-forming galaxies are distributed uniformly at 1-3r$_{200}$
 from the cluster centre, but decline steeply within r$_{200}$. The blue passive
 galaxies on the other hand show a constant distribution, except for a small
 increase around 2r$_{200}$.
 
 \item Galaxy evolution studies which base their sample selection criterion
 only on broadband colours to test Butcher-Oemler like effects, may be
 significantly affected by the presence of `complex' galaxy populations such as
 the red star-forming and blue passive galaxies discussed here. 
 \end{itemize}
 
\section{Acknowledgments}

 Funding for the Sloan Digital Sky Survey (SDSS) has been provided by
 the Alfred P. Sloan Foundation, the Participating Institutions, the
 National Aeronautics and Space Administration, the National Science
 Foundation, the U.S. Department of Energy, the Japanese
 Monbukagakusho, and the Max Planck Society. The SDSS Web site is
 http://www.sdss.org/. This research has made use of the SAO/NASA Astrophysics
 Data System. SM is supported by grants from ORSAS, UK, and the University
 of Birmingham.

 We are very thankful to the anonymous referee for the constructive comments and
 a comprehensive review of the paper. 

\label{lastpage}

%%%%%%%%%%%%%% Bibiliography %%%%%%%%%%%%%%%%%%%%%%%%%%%%%%%

\end{document}